# Non-Arrhenius conduction due to the interface-trap-induced disorder in X-doped amorphous InXZnO thin-film transistors

*Mohammed Benwadih,[1]\* J.A. Chroboczek,[2] Gérard Ghibaudo,[2]*
*Romain Coppard,[1] and Dominique Vuillaume,[3]*

[1] *CEA Grenoble/DRT/Liten, 17 rue des martyrs 38054, Grenoble, France*

[2] *IMEP-LAHC, MINATEC-INPG, 3 rue Parvis Louis Néel, 38016 Grenoble, France*

[3] *Institut for Electronics Microelectronics and Nanotechnology, CNRS, Avenue Poincaré, 59652, Villeneuve d'Ascq, France*

\*corresponding author: mohammed.benwadih@cea.fr

**ABSTRACT**

Thin film transistors, with channels composed of In-X-Zn oxides, IXZO, with X dopants: Ga, Sb, Be, Mg, Ag, Ca, Al, Ni, and Cu, were fabricated and their I-V characteristics were taken at selected temperatures in the 77K<T<300K range. The low field mobility, µ, and the interface defect density, $N_{ST}$, were extracted from the characteristics for each of the studied IXZOs. At higher T the mobility follows the Arrhenius law with an upward distortion, increasing as T was lowered, gradually transforming into the $\exp[-(T_0/T)^{1/4}]$ variation. We showed that $\mu(T, N_{ST})$ follows $\mu_0 \exp[-E_{aeff}(T,N_{ST})/kT]$, with T-dependent effective activation energy $E_{aeff}(T, N_{ST})$ accounts for the data, revealing a linear correlation between $E_{aeff}$ and $N_{ST}$ at higher T. Temperature variation of $E_{aeff}(T, N_{ST})$ was evaluated using a model assuming a random distribution of conduction mobility edge Ec values in the oxides, stemming from spatial fluctuations induced by disorder in the interface traps distribution. For a Gaussian distribution of $E_c$, the activation energy $E_{aeff}(T, N_{ST})$ varies linearly with 1/T, which accounts satisfactorily for the data obtained on all the studied IXZOs. The model also shows that $E_{aeff}(T, N_{ST})$ is a linear function of $N_{ST}$ at a fixed T, which explains the exponential decrease of µ with $N_{ST}$.

**I. INTRODUCTION**

A considerable number of papers on In-Zn oxides have been devoted to properties of these remarkable solids. They are transparent to visible light and show conductivity values of the order of 1 S.cm$^{-1}$, contradicting the paradigm of solid state physics, stating that conductive must be visible, as common metals are. Reasonably high conductivity and the detection of the field effect in the In-Zn oxides promptly led to the development of field-effect transistors with oxide channels, opening extraordinary perspectives for transparent electronics. Admixture of a fourth atom species into the In-Zn Oxides is required to assure their stability. The commonly used dopant, to borrow the term from semiconductor physics, is Ga and the resulting ternary oxide has become known as IGZO. Properties of this oxide and its applications have been discussed in numerous papers. [1,2,3] In a recent paper,[4] we have systematically explored the electrical properties of thin film transistors, TFTs, with ternary In-X-Zn oxide (IXZO) channels, where the dopants, X, were selected from a wide spectrum of atom species, namely Ga, Sb, Sn, Mg, Be, Ag, Y, Ca, Al, Ni, Cu, Mn, Mo, and Pt. The channel mobility, μ, and the associated interface defect density, $N_{ST}$, data were obtained directly from the current-voltage characteristics of the TFTs in the saturation and linear regimes.[4] Thanks to a significant number of dopants and their diversity, we were able to establish that μ and $N_{ST}$ are linked by an exponential relation,

$$\mu = \mu_0 \cdot \exp(-N_{ST}/N_{TC}), \quad (1)$$

with a universal parameter $N_{TC}$ characterizing the entire family of the TFTs used. That relation implies that the interface trap density is a determining quality factor for the ternary In-X-Zn oxide TFTs, regardless of the dopant nature.

Former studies on temperature dependence of transport properties in IGZO transistors by Lee et al.[5] that trap-limited mechanism is dominating at low gate voltages, while a percolation mechanism prevails at higher gate voltages. Kamiya et al.[1,2] carried out Hall mobility measurements on Ga-doped (IGZO) devices, made with some variations in fabrication conditions, which resulted in a certain dispersion in specimens' mobility, varying from about 10 cm$^2$/Vs to 3 cm$^2$/Vs, measured at room temperature (RT). They explained their data with a percolation model. In this work, the set of specimens we used offered a wider range of μ variations, ranging from 10 cm$^2$/Vs to about 10$^{-3}$ cm2/Vs at 300K, descending to about 10$^{-5}$ cm$^2$/Vs at the liquid nitrogen temperature. Our μ(T) data and those reported in refs [1, 2, 5] show several common features, notably a distortion of the Arrhenius plots at lower temperatures, with the limiting $\exp\{-(T_0/T)^{1/4}\}$ dependence characteristic for the variable range hopping conduction in disordered solids.[5] Note that, this mode of transport is incompatible [6] with the detection of the Hall effect in IGZO, reported in refs [1] and [2]. Here, we show that our μ(T) data, extracted from transistor current-voltage measurements at high gate voltages (around the maximum of the transconductance), are well explained by the percolation model. For every dopants used in this work, we extracted the main parameters of the gaussian distribution of activation energies (mean value of energy barrier and standard deviation). These parameters are related with the measured density of interface states, and we extracted a critical density $N_{TC}$ for the entire series of dopants, in agreement with the exponential dependance of µ with $N_{ST}$ experimentally observed in our samples.

## II. Materials and methods

The TFTs used in this work were fabricated by the Sol-Gel deposition of IXZO films on heavily doped ($p^{++}$) Si wafers, with thermally grown $SiO_2$ gate dielectric (100 nm thick). We fabricated source-drain electrodes (Ti/Au : 10nm/30nm) by evaporation, optical lithography and lift-off, giving a bottom-up TFT device structure with channel length of 20 µm and channel width of $10^4$ µm (see details in [4]). In this study we used, Ga, Sb, Be, Mg, Ag, Ca, Al, Ni, and Cu, arranged here in the order of diminishing RT mobility values, or increasing $N_{ST}$ (see section III). The sol-gel process involved the use of a precursor, such as acetate or nitrate, chloride, in an appropriate solvent salt of the metal of interest.[4] We used the same acetate-based precursor for all elements involved, for easier comparisons of properties of the synthesized ternary oxides.[4] The molar ratio of indium, X, zinc (In:X:Zn) in the precursor solution was kept constant at the 1:0.1:2 ratio. The indium concentration was maintained at 0.2M and the concentration of zinc was fixed at 0.4M. The dopants concentration was kept constant at 0.02M. The molar ratio of ethanolamine to indium and zinc was maintained at 1:8. The solution was stirred at 70°C in air for 1h and aged for 12h in air prior to the synthesis. A detailed physico-chemical analysis of the fabricated IXZO thin films is given in Ref. 4 The precursor solution of IXZO was spin coated (2000 rpm, 25s) on the above described substrates to fabricate thin-film transistors (substrates cleaned in ultrasonic acetone bath to remove the polymer protection layer, rinsed in deionized water, acetone, and isopropanol and dried). Then, the samples were annealed in air at 450° C for 10 min (hot plate) in order to decompose the precursor and form the metal oxide layer. This process was repeated three times in order to obtain the desired film thickness (about 12-15 nm).

The measurements on the TFTs were carried out in the Microtech PM150 continuous flow cryostat, under point probes, with the Agilent 5155 semi-conductor analyzer for data taking and storage. The $I_{DS}(V_G)$ characteristic in linear regime ($V_{DS}$ = 0.5 V) were taken at several temperatures in the 77K<T<300K range and the mobility values in linear regime were calculated from $I_{DS}$ versus $V_G$ plots using the so-called Y method [8] to be not perturbed by contact resistances. It implies that the mobilities are extracted at high gate voltages. The interface trap density $N_{ST}$ in Eq. (1) was obtained from the sub-threshold slope of $I_d(V_g)$ and independently from the low frequency noise (LFN) measurements.[4] We carried out LFN measurements using a home-made system in the frequency range 1 Hz - 10 kHz. We found that the noise power spectral density, $S_{ID}$, for all the devices (all dopant X) follows a 1/f law and a $I_{DS}^{\alpha}$ dependence with α about 2. These variations are in agreement with a charge density fluctuation model [7] from which with extracted $N_{ST}$ using established equations for this model. [4,8,9,10]

## III. The IXZO channel TFTs at low temperatures

Figures 1A and 1B show transfer characteristics, $I_{DS}(V_G)$ taken at various temperatures (inset) on two selected TFTs with the InSbZnO and the InNiZnO channel materials, respectively. They represent the two extreme cases in our data, of the highest and the lowest μ. As we had previously shown [4] the TFTs with Sb-dopant had μ values (≈ 8 cm$^2$/Vs) comparable with those of the IGZO channel transistors, recognized as the best. The lowest mobility of channel carriers was found in the ternary InNiZnO channel TFTs (0.8 cm$^2$/Vs) and as shown in [4] such transistors had the highest defect density ($N_{ST} \approx 2 \times 10^{13}$ eV$^{-1}$cm$^{-2}$). Nevertheless, the characteristics taken in TFTs with high defect concentrations, are seen to have correct behavior, with acceptable $I_{on}/I_{off}$ values, exceeding 10$^3$. They were stable and reproducible.

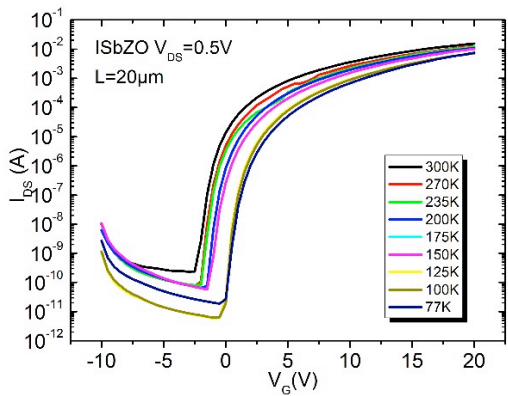
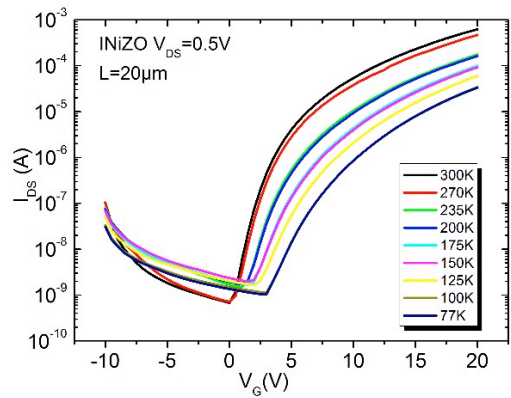

Fig. 1A. Transfer characteristics of a TFT transistor with the channel composed of InSbZnO, taken at various temperatures. Note that $I_{DS}$ attains the value of 10 mA at $V_G$=20V and the sub-threshold swing is steep.

Fig. 1B. Transfer characteristics of a TFT transistor with the channel composed of InNiZnO at various temperatures. The dispersion of the curves is higher for the Ni dopant than for the Sb dopant, being a consequence of a higher defect density in the former.

The mobility data extracted from the $I_{DS}(V_G)$ characteristics are plotted in Fig. 2A as a function of 1/T and in Fig. 2B as a function of 1/T$^{¼}$ for the series of the TFTs having channels composed of IXZO, with X dopants listed in the figure in the ascending order of $N_{ST}$.

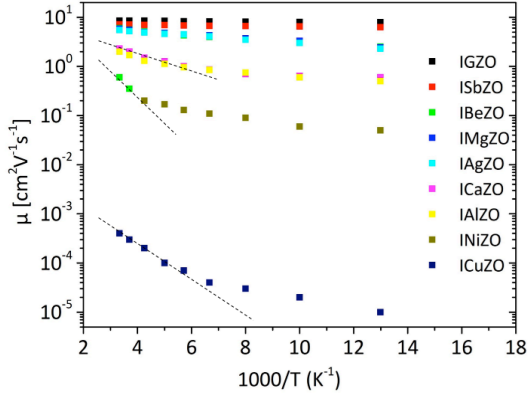 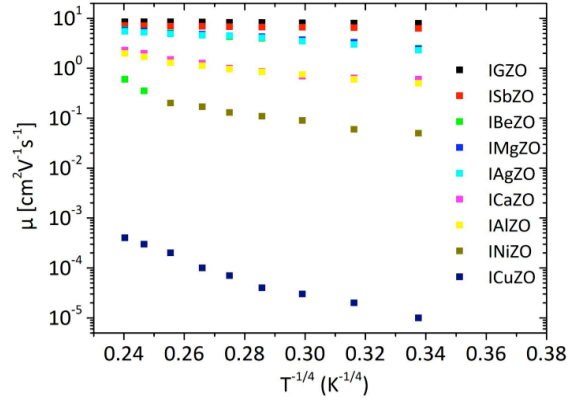

Fig. 2A. Mobility data displayed as a function of 1000/T. In the high-temperature region the curves have short linear sections, indicating mobility activation, as exp(-$E_a$/kT). The latter is seen to increase with $N_{ST}$ (the lowest in IGZO and highest in ICuZO). Dashed lines are guides for the eye.

Fig. 2B. The same data as in Fig. 2A, replotted as a function of $T^{-\frac{1}{4}}$. At lower temperatures (right-hand side of the figure) the data points are seen to follow a linear dependence.

Note that the data points in Fig. 2A follow closely the exp(-$E_a$/kT) dependence (simple activation) in the high temperature region of the plot, with an upward departure from linearity at lower temperatures. However, when the data are replotted versus $1/T^{\frac{1}{4}}$ (Fig. 2B), the plots corresponding to lower temperatures, are seen to be linear. That temperature region becomes wider in specimens having a higher density of defects (lower µ), with simultaneous shrinking of the activated transport region. Note that for samples with the lowest $N_{ST}$ (X=Ga, Sb), the mobility is not thermally activated.

Measurements of very low mobilities were possible thanks to the use of standard FET parameter extraction method from $I_{DS}(V_G)$ characteristics. It is worth noting that such a technique gives the low field µ near the MOS transistor threshold. Kamyia et al.[1,2] obtained the mobility data from Hall measurements, which probably imposed a detectability limit on their data collecting. On the other hand, the detection of the Hall effect in the low T limit, in the IGZOs provides an evidence of band transport in the In-Zn Oxides that we assumed to apply to the IXZO materials used in this work.

However, the exp{-$(T_0/T)^{\frac{1}{4}}$} mobility dependence on temperature appears also in solids where the carriers move by percolation in a system with random barriers. That was demonstrated in amorphous Si by Adler et al. [11] and more recently in certain glasses by Bischoff et. al.[12] . Kamiya et al.[1,2] adopted the model of Adler et al.[11] and showed that it accounted satisfactorily for the data. In the following, we show that the same approach applies to our data taken on the IXZOs at 77K<T<300K. In essence, the model involves an assumption that µ($N_{ST}$,T)=$µ_0$.exp{-$E_a$($N_{ST}$,T)/kT}, i.e. that the activation energy for mobility, $E_a$($N_{ST}$,T), depends on temperature. We adopt here the Adler-Kamiya transport model for interpreting our mobility data taken at various temperatures, in the TFTs having various interface trap density concentrations, originating from the diversity of the dopants species in the IXZO transistor channels.

**IV. Temperature-dependent Mobility in a Disordered System**

The existence of an activation energy at higher T suggests transport involving carrier excitation into a band of non-localized (free) states. In disordered solids, such as amorphous Si (α-Si) the bottom of the conduction band is known to fluctuate and a tail of states is formed below the conduction band, due to the presence of the disorder-generated potential fluctuations. The carriers in the tail below a certain energy $E_c$, called mobility edge, are localized in the potential wells and those for $E > E_c$ can move freely in the solid. Denoting their respective concentrations by $n_t$ and $n_f$, the conductivity, $\sigma$ can be written as,

$$\sigma = q\, n_f\, \mu_0 \qquad (2)$$

Assuming for simplicity a constant density of states and Boltzmann statistics, we can readily obtain an expression for the effective mobility,[13] $\mu$, as a function of the concentrations of the free $n_f$ and localized carriers, $n_t$ as,

$$\mu(T) = \mu_0 \frac{n_f}{n_f + n_t} = \mu_0 \exp\left(-\frac{E_c}{kT}\right) \qquad (3)$$

In a disordered system the energy $E_C$ fluctuates across the sample, entailing fluctuations in the concentration of free carriers, which translates into fluctuations in the effective mobility, by virtue of Eq. (3).

Equation (3) links the fluctuations of $\mu$ to the fluctuations of $E_c$. The problem of finding the mean value of $\mu$ is reduced now to a proper averaging of $E_c$. Following Kamiya et al.[1,2,14] and Bischoff et al.[12], we assume for simplicity that the $E_c$ distribution is Gaussian,

$$P(E_C, \alpha, \beta) = \frac{1}{\sqrt{2\pi}\beta} \exp\left(-\frac{(E_c - \alpha)^2}{2\beta^2}\right) \qquad (4)$$

where α is the mean value of $E_c$ in the distribution and β is its standard deviation.

The convolution of $P(E_c,\alpha,\beta).\mu(E_c)$ gives the most probable mobility for a given $E_c$. As electrons are locally excited to the band at various $E_c$, the mean mobility over the entire system is a sum over all available $E_c$, which can be calculated by integrating $P(E_c,\alpha,\beta).\mu(E_c)$ over $E_c$,

$$\mu(T) = \mu_0 \int_{-\infty}^{+\infty} P(E_c) \exp\left(-\frac{E_c}{kT}\right) dE_c \qquad (5)$$

The integration can be done analytically [2, 3, 12, 14] and the result is,

$$\mu(T) = \mu_0 \exp\left(\frac{\left[\alpha - \frac{q\beta^2}{2kT}\right]}{kT}\right) \qquad (6)$$

The expression in the square bracket in the exponential function argument can be considered as a T-dependent effective activation energy for the mobility,

$$E_{aeff}(T) = \alpha - \frac{q\beta^2}{2kT} \qquad (7)$$

The constant term α corresponds to the high-T limit of the activation energy, whereas the second, T-dependent term produces its decrease as the temperature is lowered, resulting in an upward distortion of the Arrhenius plot as observed in Fig. 2A.

The next step in our data analysis involved evaluating the effective activation energy for each temperature and each sample using Eqs (5-7). If we assume that $\mu_0$ in Eq. (1) is independent of temperature, we get $E_{aeff}(T, N_{ST})$=-$(kT/q).\ln(\mu(T,N_{ST})/\mu_0)$. By this way, a continuous variation of $E_{aeff}(T,N_{ST})$ can be obtained as a function of temperature and $N_{ST}$ for all the samples as can be seen in Fig. 3A. The temperature dependence of $E_{aeff}$ (Fig. 3A) can be well fitted (solid lines) by the Adler-Kamiya model of Eq. (6) with best fit α and β parameters shown in Fig. 4. It should be noted that the $E_{aeff}(T)$ plots flatten out near 300K, which is consistent with a temperature-independent $E_{aeff}$ at sufficiently elevated T where α term dominates. Moreover, as indicated by Fig. 3B, $E_{aeff}$ at room temperature varies almost linearly with the trap density $N_{ST}$ as $E_{aeff}$=K.$N_{ST}$. Therefore, the latter well accounts for the exponential decrease of μ with the oxide trap density $N_{ST}$ with $N_{TC}$=kT/K in Eq. (1). From Fig. 3B, K is about $8\times10^{-15}$ eV$^2$cm$^{-2}$, thus we have $N_{TC} \sim 3\times10^{12}$ eV$^{-1}$cm$^{-2}$ at room temperature in good agreement with our previous determination from RT measurements.[4]

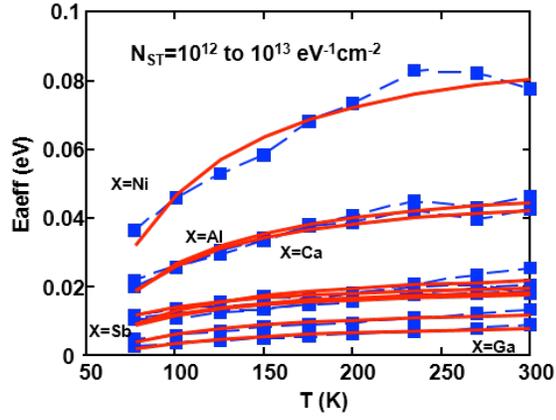 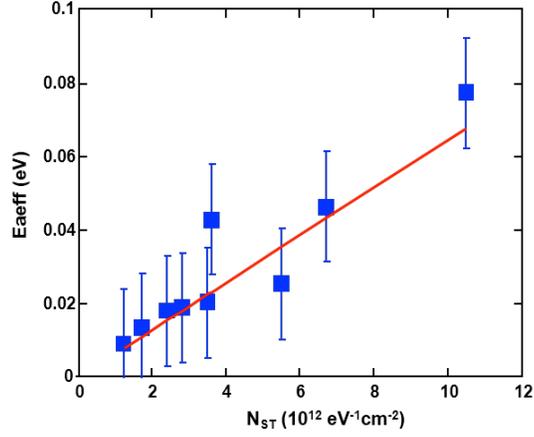

Fig. 3A. Effective activation energy $E_{aeff}$ for mobility as a function of T (squares: data points, blue on line) in IXZO channel TFTs doped with nine different X atom species (list in the text). Solid line curves represent calculation results involving Eq. (6) and (7).

Fig. 3B. Effective activation energy $E_{aeff}$ for mobility as a function of $N_{ST}$ at 300K. $E_{aeff}$ is found to vary linearly with $N_{ST}$.

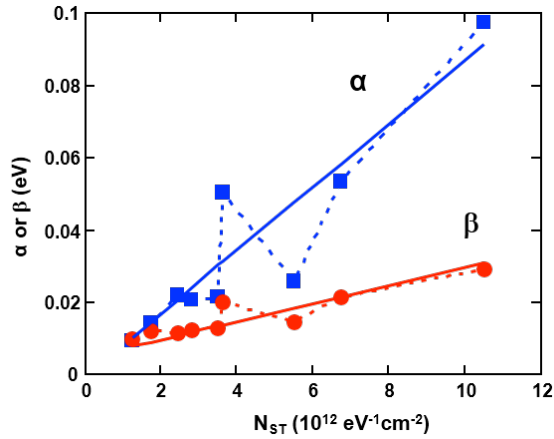

Fig. 4. Variation of α and β parameters with trap density $N_{ST}$ for various IXZO TFTs. The lines serve to guide the eye.

In Fig. 4 the values of the parameters α and β in Eq. (6) and (7), of the Gaussian distribution of the activation energies values for mobility fluctuations, are plotted as a function of $N_{ST}$, for all the transistors we used. As seen both α and β vary linearly with $N_{ST}$. That means that at a higher trap density the disorder at channel-dielectric interface is higher, as expected. That corroborates the conclusion of our former paper stating that the mobility in the oxide channels is principally determined by the defects density and not by their chemical characteristics (e.g. ionic radius, electronegativity).

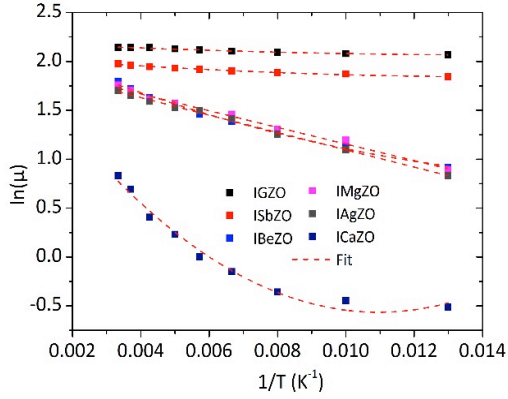 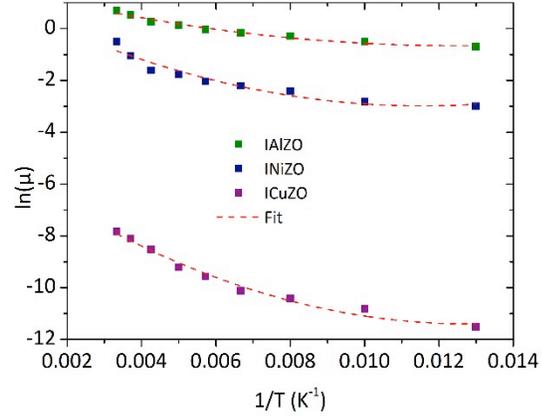

Fig. 5A. Mobility data (points) for six different dopants (listed in the drawing) in IXZO are seen to be well accounted for by the calculation, involving Eqs (5 and 6) with appropriate choice of parameters α and (Fig. 4) Red (on line) dash curves.

Fig. 5B. Mobility data and the results of the calculation for the three dopants giving lower μ values in the oxides, than those in Fig. 5A.

Figures 5A and 5B, show that the mobility variation with T for the entire set of transistors with various composition of the oxide channels is well accounted for by calculation involving Eq. (6) with the parameters α and β optimized for each dopant, so as the best fit to the data is obtained. This result shows that the transport model discussed above is perfectly adapted for the IXZO transistors.

## V. CONCLUSIONS

We present in this paper results on mobility measurements at the temperatures varied in the 77K<T<300K range, carried out on a set of nine ternary InXZn oxides doped with different X atom species. The diversity of the dopants we used resulted in a wide variation of the defect density in the devices and gave mobility dispersion much wider than that reported in the study of Kamiya et al.[1,2] and Lee et al.[5] Our data show that the model of transport assuming a Gaussian distribution of activation energies for mobility, accounts for the μ(T) data obtained on the entire set of thin film transistors with the IXZO channels of varied composition. Our results also show that the effective activation energy for the mobility varies with reciprocal temperature according to the predictions of the Adler-Kamiya model. In particular, the mean values and the standard deviation coefficients in the activation energy values for the transistors are shown to be proportional to the trap density at the transistor channel-dielectric interfaces. That means that at higher trap density the disorder becomes more pronounced, which, in turn, enhances the non-Arrhenius behavior in electron transport in the IXZO transistor channels.

**REFERENCES**

[1] T. Kamiya, K. Nomura, and H. Hosono, "Origin of definite Hall voltage and positive slope in mobility-donor density relation in disordered oxide semiconductors" Appl. Phys. Lett., **96**, 122103, 2010.

[2] T. Kamiya, K. Nomura, and H. Hosono," Electronic Structures Above Mobility Edges in Crystalline and Amorphous In-Ga-Zn-O: Percolation Conduction Examined by Analytical Model" J. Disp. Techn, **5**, 462, 2009.

[3] S. Lee, K. Ghaffarzadek, A. Nathan, J. Robertson, S. Jeon, C.-J. Kim, I-H. Song, and U-I. Chung," Trap-limited and percolation conduction mechanisms in amorphous oxide semiconductor thin film transistors" Appl. Phys. Lett. **98**, 203508, 2011.

[4] M. Benwadih, J.A. Chroboczek, G. Ghibaudo. R. Coppard, and D. Vuillaume," Impact of dopant species on the interfacial trap density and mobility in amorphous In-X-Zn-O solution-processed thin-film transistors" J. App. Phys., **115**, 214501, 2014.

[5] S. Lee, A. Nathan, J. Robertson, K. Ghaffarzadeh, M. Pepper, S. Jeon, C. Kim, I-H. Song and K. Kim,"Temperature dependent electron transport in amorphous oxide semiconductor thin film transistors" Int. Electron Dev. Meeting, 14.6.1, 2011.

[6] N.F. Mott, Phil. Mag., **19**, 853 (1972).

[7] G. Ghibaudo. Electron. Lett. 24, 543 (1998)

[8] B. A. L. McWorther, Semiconductor Surface Physics, edited by R. H. Kingston University of Pennsylvania, Philadelphia, PA, p. 207, 1957.

[9] J. A. Chroboczek, IEEE ICMTS. 03CH37417, 95 (2003).

[10] G. Ghibaudo, Microelectron. Eng. 39, 31 (1997).

[11] D. Adler, L.P. Flora, S.D. Senturia," Electrical conductivity in disordered systems" Solid State Commun., **12**, 9, 1973.

[12] C. Bischoff, K. Schuler. S. P. Beckman, and S. W. Martin," Non-Arrhenius Ionic Conductivities in Glasses due to a Distribution of Activation Energies" Phys. Rev. Lett., **109**, 07590, 2012.

[13] J.L. Robert, B. Pistoulet, A. Reymond, J.M. Dusseau, and G.M. Martin," New model of conduction mechanism in semi-insulating GaAs" J. App. Phys., **50,** 349, 1979.

[14] M. Kimura, T. Kamiya, T. Nakanishi, K. Nomura, and H. Hosono," Intrinsic carrier mobility in amorphous In–Ga–Zn–O thin-film transistors determined by combined field-effect technique" Appl. Phys. Lett. **96**, 262105, 2010.